# Influence of Alcohol Smell and Imagination on the Condition of the Human Organism and Subjective Human Experience

Tatiana Berezina

## Abstract

In this study, alcohol smell and imagination of alcohol has been shown to change the condition of a human organism. In this study, alcohol smell has been shown to change the condition of a human organism. Based on the test results, the presence of alcohol was observed upon breathing out, but was rarely observed in spits and much rarer in urines. This effect is most evident shortly after an olfactory perception of alcohol and continues for 60 min for some test persons. The test persons also noted the appearance of a subjective feeling of alcohol intoxication. However, the condition developed when the alcohol smell perception differed from the classical alcohol intoxication, i.e., the test subjects only have a few intoxication symptoms, and not one person would have all of them.

**Key words:** alcohol, olfactory perception, imagination, personality, and intoxication

## Introduction

Smells and Imagination can influence a person's psychological and physiological condition. Olfaction is known to play a key role in the synchronization of women's menstrual periods, called the McClintock effect (Russell *et al.*, 1980). It was also proved that air aromatization, including the introduction of volatile steroid compounds in air (steroid is a component of men's sweat), can decrease pain perception in some categories of test persons (Villemure and Bushnell, 2007). Natural pleasant smells have been observed to improve people's mood and decrease their level of alertness (Weber and Heuberger, 2008). The fragrance of brandy mint improves a person's cognitive skills (Barker *et al.*, 2003). Smells are related to human memory and can refresh memory, i.e., a person will get the same emotions experienced at the moment of that smell perception (Herz, 2005). Alcohol smell can similarly influence the functioning of the human organism. For instance, it affects sensitivity to other smells (Patel *et al.*, 2004) and their subjective preference (Mennella, 2007).

As well as the smell, imagination of alcohol influences force of influence on a functional condition of an organism. That imagined chemical substances can cause in the person symptoms corresponding to their application if he trusts in their reality, in medicine well-known and carries the Effect of a placebo. On healthy volunteers by a method of classical conditioned reflexes it has been established that usual soft drink if to tell that this alcohol, the same effect, as a unitary dose of alcohol caused; at women - authentic increase of a rhythm of heart, and at men – a bradycardia (Newlin, 1989).

## Materials and Methods

Test persons included 100 young adults (including 50 males). Experimental bunch of 50 persons: 25 men and 25 women. Control bunch – 50 persons: 25 men and 25 women. Examinees were students of the correspondence unit, the studying humanities, at the age of 20 - 45 years.

**Methods**

1. Drivesafe alcohol detection device is used to measure ethyl alcohol concentration while breathing out. The measurement range is 0 –1.5 pm (promille). 1 promille - 1 gramme of alcohol on 1 litre of blood. 1 promille = 0,1 BAC. The device carries compliance and registration certificates from the Ministry of Health Care and Social Development of the Russian Federation. All medical examinations related to this study were then held within 3 months after calibration.

2. Reagent strips for the qualitative detection of alcohol in spit: SIMS-2 alcohol sensor. Grades: traces (+) – a strip turns green, but a color degree does not attain the least value of the quantified values; 0.2 pm, 0.5 pm.

3. Reagent strips for the qualitative detection of alcohol in urine: SIMS-2 alcohol sensor. Grades: traces (+) – a strip turns green, but a color degree does not attain the least value of the quantified values; 0.2 pm, 0.5 pm.

4. Container with alcohol. 100 ml of 95% ethyl alcohol poured into 0.5 L cola bottle.

5. Rotative test (Taschen, 1955). Estimation of the degree of alcohol intoxication by the presence of ocular nystagmus. A test person is rotated five times through a full 360° during 10 s, and then he is asked not to turn his head or close his eyes. Afterwards, the test person looks at the object placed in front of him, and the experimenter measures the duration of nystagmus (high or mean amplitudes) using a stopwatch timer.
The duration of nystagmus measuring 8 s points to a very light degree of alcohol intoxication; 9–14 s, light degree of alcohol intoxication; and 15 s, high degree of alcohol intoxication.

6. Method of self-estimation of the organism's condition (original scale). A test person was asked to estimate subjectively his condition by the following scale:
Absolutely sober – 1 point

Rather sober than drunk – 2 points

Slightly drunk – 3 points

Significant alcohol intoxication – 4 points

Very drunk – 5 points

**Procedure**

**Experimental bunch**.

1. Preliminary diagnostics.

    a) A test person was subjected to measurement of ethyl alcohol while exhaling using the Drivesafe alcohol detection device. Three measurements were carried out.

    b) Organism condition was self-estimated.

    Only test persons considering themselves to be absolutely sober and having 0 pm in breathing out in all three tests, participated in further experiments.

2. Experimental treatment was carried out in a game format. We gave a bottle with alcohol to the test persons and offered to determine the ethyl alcohol content in the drink. Two to four persons usually participated in the test, and we added an element of competition: "Let's check who has better intuition and cognitive abilities!" Using that method, we wanted to increase the motivation of the tests persons to smell alcohol. A test person smelt the alcohol in a bottle, suggested the ethyl alcohol concentration, and then breathed out into the Drivesafe alcohol detection device to determine the actual ethyl alcohol concentration. The person was tested five times in the same sequence within 5 min.

3. Results of the experiment

Several series were carried out.

    a) 1st series – Estimation of smell influence 15 min after the test started. Within 10 min, a test person stayed in a room, spoke with other people and did not have any contact with alcohol. The experimenter sustained a conversation with the test persons about the

experiment carried out and about alcohol in general. The experimenter then measured the presence of ethyl alcohol as the test persons breathed out using the Drivesafe alcohol detection device. This was repeated five times.

b) 2nd series – Measurement of smell influence 30 min after exposure. Analogous.

c) 3rd series –Measurement of smell influence 60 min after exposure. Analogous.

The subjective human condition was done by the self-estimation scale.

The presence of ocular nystagmus was estimated by means of the rotative test.

The presence of alcohol in spit was estimated.

The presence of alcohol in urine was estimated.

**Control bunch.**

Examinees were offered to track, as varies at them their usual functional state during a half an hour without any influence. Experimental influence on them it was not carried out (they did not nose alcohol and expressly did not imagine it).

1) Organism condition was self-estimated.

2) 1st series. A test person was subjected to measurement of ethyl alcohol while exhaling using the Drivesafe alcohol detection device. Five measurements were carried out.

3) 2st series – Estimation of smell influence 15 min after the test started.

4) 3nd series – Measurement of smell influence 30 min after exposure. Analogous.

5) 4rd series –Measurement of smell influence 60 min after exposure. Analogous

6) The subjective human condition was done by the self-estimation scale.

7) The presence of alcohol in spit was estimated.

8) The presence of alcohol in urine was estimated.

**Data Analyses**

We formed a "variable objects" matrix used as variable parameters:

1) total alcohol content while breathing out right after olfactory exposure;
2) total alcohol content while breathing out 15 min after olfactory exposure;
3) total alcohol content while breathing out 30 min after olfactory exposure;
4) total alcohol content while breathing out 60 min after olfactory exposure;
5) alcohol content in spit;
6) alcohol content in urine; and,
7) self-estimation of intoxication level.

We used the STATISTICA 6 program and carried out a correlation analysis to detect any relationships between variable parameters. Marked correlations are significant at $p < 0.05$, $N=50$. We used the STATISTICA 6 program and carried out a one-way ANOVA to analyze designs with a single categorical independent variable (factor). Marked correlations are significant at $p < 0.05$, $N=50$.

Dependent variables

1. total alcohol content while breathing out right after olfactory exposure;
2. total alcohol content while breathing out 15 min after olfactory exposure;
3. total alcohol content while breathing out 30 min after olfactory exposure;
4. total alcohol content while breathing out 60 min after olfactory exposure;
5. alcohol content in spit;
6. alcohol content in urine; and,
7. self-estimation of intoxication level.

Categorical factor: Alcohol odour. 1 level – the examinee did not nose alcohol, 2 level the-examinee nosed alcohol.

## Results

**Alcohol while breathing out:**

The results of the change in breath composition after alcohol smell influence are listed in Table 1. The results show that after exposure to the smell of alcohol, 42% of the test persons had alcohol while breathing out. Moreover, 14% had alcohol concentrations exceeding 5 pm while breathing out, 15 min after exposure. On the other hand, 16% had alcohol while breathing out, of which 2% had alcohol concentrations exceeding 5 pm, 30 min after exposure, 16% of the persons still had alcohol while breathing out, but not > 0.5 pm. An hour later, 8% of the test persons had alcohol while breathing out, but not > 0.5 pm

**Alcohol in spit and urine**:

The results of alcohol content in spit and urine 1 h after alcohol smell influence are given in Table 2. The results show that 28% of the test persons had alcohol in spit; 18% had only traces of alcohol, 8% had 0.2 pm of alcohol, and 2% had 0.5 pm of alcohol. In 2% of the test persons, 0.2 pm of alcohol was found in their urine.

**Presence of ocular nystagmus:**

All the test persons subjected to alcohol smell did not have ocular nystagmus 1 h after the experiment.

**Subjective feeling of alcohol intoxication**:

The results of the test person's self-estimation of their condition are listed in Table 3. It was found that 46% of the test persons considered themselves absolutely sober, while 54% felt that their condition changed; 28% were unsure about their condition (sober than drunk), 20% reported that they were slightly drunk, and only 6% felt that they were significantly drunk.

**Interrelation of indicators:**

Most characteristics of the condition of the human organisms are interrelated (Table 4). The alcohol level registered shortly after the olfactory exposure correlates with the alcohol level registered while breathing out 15 and 30 min after the smell perception, as well as with the presence of alcohol in spit and with the subjective feeling of alcohol intoxication. However, this indicator does not correlate with the appearance of alcohol in urine and alcohol level while breathing out 60 min after the smell perception. Despite this, we observed a gradual change in the condition of the organism. 60 min after alcohol smell perception. The alcohol level while breathing out correlates with the appearance of alcohol in spit and urine, but is not related to the subjective feeling of alcohol intoxication.

**Results of control bunch**.

In the absence of influence within an hour no changes of a functional state descended. Alcohol has never occurred, neither in an expiration of examinees, nor in a saliva, in urine. However there was a minor alteration of value judgment by examinees of the state. 92 % of a h continued to consider themselves "absolutely sober", and 8 % have stated doubt in it, having chosen the answer «is faster sober».

**One-way ANOVA**.

We estimated reliability of influence of an odour with the help one-way ANOVA ([Table 5](#)). The alcohol odour authentically influences alcohol appearances in an expiration after perceptions. The alcohol odour authentically influences alcohol appearance in a saliva in an hour after perception. The alcohol odour authentically influences on Self-estimation. The alcohol odour does not influence alcohol appearance in an expiration in 15-60 minutes after perception (one-way ANOVA has shown absence of effect). However in a reality alcohol in an expiration occurred. We explain it influence of any other factor – not an odour. We assume, that it can be effect of imagination. Our assumption demands additional research.

# Discussion

Our tests show that the condition of the organism changes under the influence of alcohol smell. It ought to be noted that endogenous ethanol is a natural metabolite of an organism and is present in human and animal blood in insignificant concentrations (up to 0.15 mg %) (Walker and Curry,

1966). The endogenous ethanol level varies and can increase or decrease under the influence of many factors (Lester, 1962), for instance, as a result of autohypnosis. The placebo effect is characteristic of alcohol. For example, it was revealed that an ordinary cold drink (e.g., 7 Up) can cause symptoms of alcohol intoxication, if the test persons are told that the drink is alcoholic. As a result, the cold drink causes the same effect as a single dose of alcohol: women suffer from acceleration of heart rhythm, and men, from brachycardia (Newlin, 1989). In our tests, we combined autohypnosis with alcohol smell exposure. The memory of smells is known to be the strongest and is related to the emotional condition of a person (Herz et al., 2004). Under the influence of an aroma, a person not only remembers the situation related to that smell, but also the emotional condition when the smell was perceived; in our case, the symptoms of alcohol intoxication can be reproduced. However, in our opinion, the reproduced condition of alcohol intoxication is not classical, because a test person only presents a few alcohol intoxication symptoms. In the case of real alcohol consumption, a drunken person will have all the symptoms. Most of the test persons had alcohol while breathing out but did not have it in spit or urine. In such a case, the Drivesafe alcohol detection device is likely to sense not only ethyl alcohol, but any of the other metabolites of the organism (aldehydes, alcohols). These metabolites appear while breathing out in response to alcohol smell exposure and were absent prior to exposure. However, some test persons showed several symptoms of alcohol intoxication.

## Summary and Conclusions

Alcohol smell and imagination of alcohol influences the human organism's condition. Shortly after alcohol smell perception, the alcohol content registered by the Drivesafe alcohol detection device appears while breathing out in 42% of the test persons; in 16% after 15 and 30 min; and in only 8% after 60 min. Alcohol smell exposure also results in the appearance of alcohol in spit among 28% of the persons and in the urine of 2%. Influence of an odour of alcohol is significant at once while breathing, in spit and self-estimation ($p < 0.05$). Most of the characteristics of the human organism's condition are interrelated. However, there is no absolute correlation between the alcohol olfactory perception, its presence while breathing out some time later, its appearance in the spit and urine, and the subjective feeling of alcohol intoxication. Some test persons had alcohol while breathing out 1 h after smelling the alcohol, but did not have it in their spit and urine, while other persons virtually did not have alcohol while breathing out, but had it in their spit. Some persons subjectively felt drunk, but neither had alcohol while breathing out nor in the spit and urine. We think that the condition, which develops after the alcohol smell perception, is similar to classical alcohol intoxication, but does not correspond to it. This was proven by the fact that the test persons did not have ocular nystagmus after smelling the alcohol. Thus, alcohol smell does not influence the human organism's equilibrium system and does not cause the classical symptoms of alcohol intoxication.

# References

Barker S, Grayhem P, Koon J, Perkins J, Whalen A, Raudenbush B. Improved performance on clerical tasks associated with administration of peppermint odor. Perceptual and Motor Skills, (2003) 97, 1007–10.

Herz RS, Eliassen J, Beland S, Souza T. Neuroimaging evidence for the emotional potency of odor-evoked memory. Neuropsychologia, (2004) 42, 371–8.

Herz RS. Odor-associative learning and emotion: effects on perception and behavior. Chemical Senses, (2005) 30 Suppl 1, i250-51.

Lester D. The concentration of apparent endogenous ethanol. // Quart. J. Stud. Alcohol., 1962, vol. 23, p. 17–26.

Mennella J.A. The chemical senses and the development of flavor preferences in humans // in Textbook on Human Lactation, edited by P. E. Hartmann and T. Hale (Texas: Pharmasoft Publishing, 2007).

Newlin D. B. Placebo responding in the same direction as alcohol in women // Alcoholism: Clinical and Experimntal Research. 1989, v.13, № 1, p.36-39.

Newlin D. B. Placebo responding in the same direction as alcohol in women // Alcoholism: Clinical and Experimental Research. 1989, v.13, № 1, p.36–39.

Patel SJ, Bollhoefer AD, Doty RL .Influences of ethanol ingestion on olfactory function in humans. Psychopharmacology, (2004) 171, 429-34.

Russell M.J., Switz G.M., Thompson K. Olfactory influence on the human menstrual cycle. Pharmacology, Biochemistry and Behavior, 1980,13, 737–738.

Taschen B. Eine einfache Nystagmoprobe zur Fiststellung der Alkoholbeenflussung. Med. Wschr., 1955, I.25.

Villemure C, Bushnell MC. The effects of the steroid androstadienone and pleasant odorants on the mood and pain perception of men and women. European Journal of Pain, (2007) 11, 181 - 191.

Walker G. W., Curry A. S. "Endogenous" alcohol in body fluids // Nature (London), 1966, vol. 210, p. 1368.

Weber S, Heuberger E. The Impact of Natural Odors on Affective States in Humans // Chemical Senses, 2008, 33(5), 441–447.

Table 1. Dynamics of alcohol in breathing out after alcohol smell perception, % of total number of test persons (right after, and 15, 30, and 60 min after smelling alcohol).

|                  | Right after | 15 min | 30 min | 60 min |
|------------------|-------------|--------|--------|--------|
| 0                | 58          | 84     | 84     | 92     |
| To 0.5 pm        | 28          | 14     | 16     | 8      |
| 0.5 pm and higher| 14          | 2      | 0      | 0      |

Table 2. Presence of alcohol in spit and urine 1 h after alcohol smell perception, % of total number of test persons.

|        | In spit | In urine |
|--------|---------|----------|
| 0      | 72      | 98       |
| Traces | 18      | 0        |
| 0.2 pm | 8       | 2        |
| 0.5 pm | 2       | 0        |

Table 3. Subjective estimation of their condition by test persons, % of total number of test persons.

|  | Absolutely sober, 1 point | Rather sober than drunk, 2 points | Slightly drunk, 3 point | Significant alcohol intoxication, 4 points | Very drunk, 5 points |
|--|---|---|---|---|---|
|  | 46 | 28 | 20 | 6 | 0 |

Table 4. Coefficients of correlation between the indicators of organism condition

|             | 15 min | 30 min | 60 min | Spit  | Urine  | Self-estimation |
|-------------|--------|--------|--------|-------|--------|-----------------|
| Right after | 0.61*  | 0.47*  | 0.11   | 0.42* | -0.09  | 0.38*           |
| 15 min      | -      | 0.71*  | 0.39*  | 0.49* | 0.02   | 0.08            |
| 30 min      | -      | -      | 0.79*  | 0.65* | 0.31*  | 0.07            |
| 60 min      | -      | -      | -      | 0.61* | 0.63*  | -0.05           |
| Spit        | -      | -      | -      | -     | 0.35*  | 0.04            |
| Urine       | -      | -      | -      | -     | -      | -0.02           |

- correlation was not calculated

* - correlation is significant at $p < 0.05$

Table 5. Assessment of reliability of influence of an odour of alcohol on a functional state (one-way ANOVA).

|  |  | Alcohol |  |  |  |  |  |  |
|--|--|---------|--|--|--|--|--|--|
|  |  | At once | 15 min | 30 min | 60 min | Spit | Urine | Self-estimation |
| influence of an odour | F | **30,6293*** | 0,13829 | 0,13440 | 0,12678 | **7,12835*** | 3,51933 | **6,0021*** |
|  | p | 0,00000 | 0,71039 | 0,71430 | 0,72646 | 0,00822 | 0,06214 | 0,01546 |

Influence is significant at $p < 0.05$

Right after – indicator of alcohol while breathing out right after alcohol smell exposure

At once - indicator of alcohol while breathing at once after alcohol smell exposure

15 min – indicator of alcohol while breathing out 15 min after alcohol smell exposure

30 min – indicator of alcohol while breathing out 30 min after alcohol smell exposure

60 min – indicator of alcohol while breathing out 60 min after alcohol smell exposure

Spit – indicator of alcohol in spit

Urine – indicator of alcohol in urine

Self-estimation – self-estimation of alcohol intoxication level.